\documentstyle[12pt]{article}

\makeatletter
\def\vereq#1#2{\lower3pt\vbox{\baselineskip1.5pt \lineskip1.5pt
\ialign{$\m@th#1\hfill##\hfil$\crcr#2\crcr\sim\crcr}}}
\makeatother

\def\lesssim{\mathrel{\mathpalette\vereq<}}

\makeatletter
\def\lesssim{\mathrel{\mathpalette\vereq<}}
\def\vereq#1#2{\lower3pt\vbox{\baselineskip1.5pt \lineskip1.5pt
\ialign{$\m@th#1\hfill##\hfil$\crcr#2\crcr\sim\crcr}}}

\makeatother

\newcommand{\beq}{\begin{equation}}
\newcommand{\eeq}{\end{equation}}

\newcommand{\remove}[1]{}

\begin{document}
\input psfig
\begin{titlepage}
\begin{center}
\today     \hfill    LBL-37432\\

{\large \bf Light U(1) Gauge Boson \\
Coupled to Baryon Number}\footnote{This
work was supported by the Director, Office of Energy Research, Office of
High Energy and Nuclear Physics, Division of High Energy Physics of the
U.S. Department of Energy under Contract DE-AC03-76SF00098.}


\vskip 0.3in

Christopher D. Carone\footnote{Talk presented by C.~Carone at the
Workshop on Particle Theory and Phenomenology, May 17-19, 1995,
Iowa State University, Ames Iowa.} and Hitoshi Murayama

{\em Theoretical Physics Group\\
     Lawrence Berkeley Laboratory\\
     University of California\\
     Berkeley, California 94720}

\end{center}

\vskip .3in

\begin{abstract}
We discuss the phenomenology of a light U(1) gauge
boson, $\gamma_B$, that couples only to baryon number.  Gauging baryon
number at high energies can prevent dangerous baryon-number violating
operators that may be generated by Planck scale physics.  However, we
assume at low energies that the new U(1) gauge symmetry is spontaneously
broken and that the $\gamma_B$ mass $m_B$ is smaller than $m_Z$.  We show for
$m_\Upsilon<m_B<m_Z$ that the $\gamma_B$ coupling $\alpha_B$ can be as
large as $\sim 0.1$ without conflicting with the current experimental
constraints.  We argue that $\alpha_B\sim 0.1$ is large enough to produce
visible collider signatures and that evidence for the $\gamma_B$ could
be hidden in existing LEP data.  We show that there are realistic models
in which mixing between the $\gamma_B$ and the electroweak gauge bosons
occurs only as a radiative effect and does not lead to conflict with
precision electroweak measurements.  Such mixing may nevertheless provide
a leptonic signal for models of this type at an upgraded Tevatron.
\end{abstract}

\end{titlepage}
\renewcommand{\thepage}{\roman{page}}
\setcounter{page}{2}
\mbox{ }

\vskip 1in

\begin{center}
{\bf Disclaimer}
\end{center}

\vskip .2in

\begin{scriptsize}
\begin{quotation}
This document was prepared as an account of work sponsored by the United
States Government. While this document is believed to contain correct
information, neither the United States Government nor any agency
thereof, nor The Regents of the University of California, nor any of their
employees, makes any warranty, express or implied, or assumes any legal
liability or responsibility for the accuracy, completeness, or usefulness
of any information, apparatus, product, or process disclosed, or represents
that its use would not infringe privately owned rights.  Reference herein
to any specific commercial products process, or service by its trade name,
trademark, manufacturer, or otherwise, does not necessarily constitute or
imply its endorsement, recommendation, or favoring by the United States
Government or any agency thereof, or The Regents of the University of
California.  The views and opinions of authors expressed herein do not
necessarily state or reflect those of the United States Government or any
agency thereof, or The Regents of the University of California.
\end{quotation}
\end{scriptsize}

\vskip 2in

\begin{center}
\begin{small}
{\it Lawrence Berkeley Laboratory is an equal opportunity employer.}
\end{small}
\end{center}

\newpage
\renewcommand{\thepage}{\arabic{page}}
\setcounter{page}{1}
\section{Introduction}
The standard model possesses a number of global U(1) symmetries, namely baryon
number, and three types of lepton number.  It has been argued, however, that
global symmetries should be broken by quantum gravity effects \cite{gravity},
with potentially disastrous consequences.  Baryon number-violating operators
generated at the Planck scale can lead to an unacceptably large proton decay
rate, especially in some supersymmetric theories \cite{proton}.  This problem
can be avoided naturally if baryon number is taken instead to be a local
symmetry at high energies.

In this talk, we consider the consequences of gauging the symmetry generated by
baryon number \cite{pwork}, U(1)$_B$.  We assume that the symmetry is
spontaneously broken and that the corresponding gauge boson $\gamma_B$ develops
a mass $m_B < m_Z$. Additional electroweak scale fermions are necessary to
render the model anomaly free, and a new Higgs field with baryon number $B_H$
is required for spontaneous symmetry breakdown.  However, by taking $B_H$ to
be small, we can raise the baryon number Higgs mass and effectively decouple it
from the problem.  Instead we will focus on the $\gamma_B$
phenomenology \cite{nt,carmur,db}, which can be described in terms of the
parameter space $m_B$-$\alpha_B$-$c$. Here $4\pi\alpha_B$ is the squared
gauge coupling\footnote{U(1)$_B$ is normalized such that
the $(\gamma_B)^\mu \overline{q}\gamma_\mu q$ coupling is
$\sqrt{4\pi\alpha_B}/3$}, and $c$ is a parameter that describes the
kinetic mixing between baryon number and ordinary hypercharge:
\beq
{\cal L}_{kin}=-\frac{1}{4} \left( F^{\mu\nu}_Y F_{\mu\nu}^Y
+ 2 c\, F^{\mu\nu}_B F_{\mu\nu}^Y + F^{\mu\nu}_B F_{\mu\nu}^B
\right) .
\label{eq:mixdef}
\eeq
Since $c$ is quite small, as we will see in Section~3, the
discussion of jet physics can be described in terms of
an effective parameter space, the $m_B$-$\alpha_B$ plane.
We show for $m_\Upsilon<m_B<m_Z$ that the $\gamma_B$ coupling $\alpha_B$
can be as large as $\sim 0.1$ without conflicting with the current
experimental constraints. In this case, a signal might be discerned by
reanalysis of existing accelerator data.

\section{Parameter Space}
Aside from the mixing effects discussed in the next section, the $\gamma_B$
boson couples only to quarks, so that its most important effects can be
expected in the same processes used in measuring the QCD
coupling $\alpha_s$.  Thus, we will determine the allowed regions of
the $m_B$-$\alpha_B$ plane by considering the following observables:

{\em $R_Z$.}  In the absence of mixing ($c \approx 0$), the $\gamma_B$ boson
contributes to $R_Z = \Gamma(Z\rightarrow {\rm hadrons})/\Gamma
(Z\rightarrow \mu^+\mu^-)$ at order $\alpha_B$ through (i) direct production
$Z\rightarrow \overline{q}q\gamma_B$, and (ii) the $Z\overline{q}q$ vertex
correction.  We require that the resulting shift in the value of
$\alpha_s(m_Z)$ away from the world average \cite{PDG} to
be within two standard deviations of the value measured at LEP,
$\alpha_s(m_Z) = 0.124\pm 0.007$ \cite{PDG}.  As shown in Fig. 1, this
roughly excludes the region of parameter space above $\alpha_B \approx 0.3$.
\begin{figure}
\vspace*{13pt}
\centerline{\psfig{file=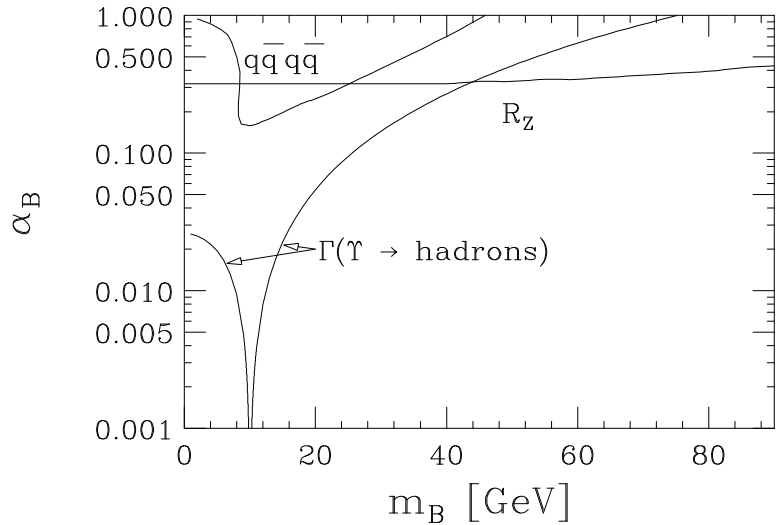,width=4in,angle=0}}
\caption{Allowed regions of the $m_B$-$\alpha_B$ plane, for $c=0$.
The region above each line is excluded.}
\end{figure}

{\em Z $\rightarrow$ jets.} The $\gamma_B$ boson contributes to $Z$ decay
to four jets, via $Z\rightarrow \overline{q}q\gamma_B$,
$\gamma_B\rightarrow \overline{q}{q}$. In doing our parton-level jet
analysis, we adopt the JADE algorithm. The four-jet cross section is shown
in Fig.~2a as a function of $y_{{\rm cut}}$, normalized to the lowest order
two-jet cross section $\sigma_0$, for $\alpha_B=0.1$ and for a range of
$m_B$ \cite{BASES}.  If we require that the fraction of four-jet events
that are four-quark jet events be less than 9.1\% (95\% C.L.)
at $y_{{\rm cut}}=0.01$ \cite{OPAL}, we exclude the region at
the top of Fig.~1.  This is only an approximate bound illustrating the
region that may be excluded by a rigorous treatment of the angular
distributions of the four-jet events involving $\gamma_B$ exchange.
Reanalysis of existing LEP data \cite{other} using a larger
value of $y_{{\rm cut}}$ and taking into account the angular distribution
expected for a massive intermediate state would be necessary
before we can put further constraints on the $m_B$-$\alpha_B$ plane.
Thus, there is also the potential of finding a signal by reanalysis
of existing data.
\begin{figure}
\vspace*{13pt}
\centerline{\psfig{file=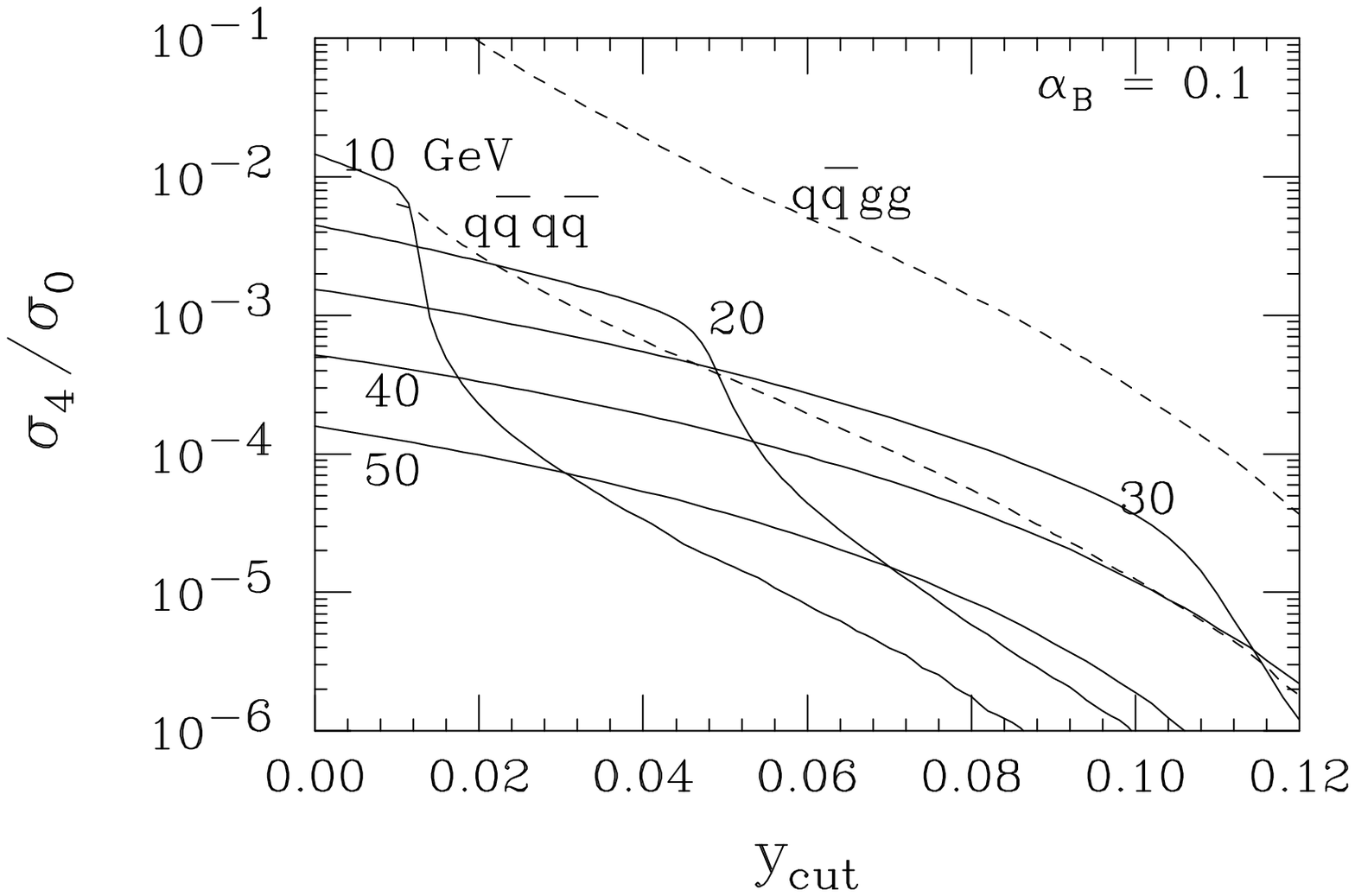,width=2.75in,angle=0}
\psfig{file=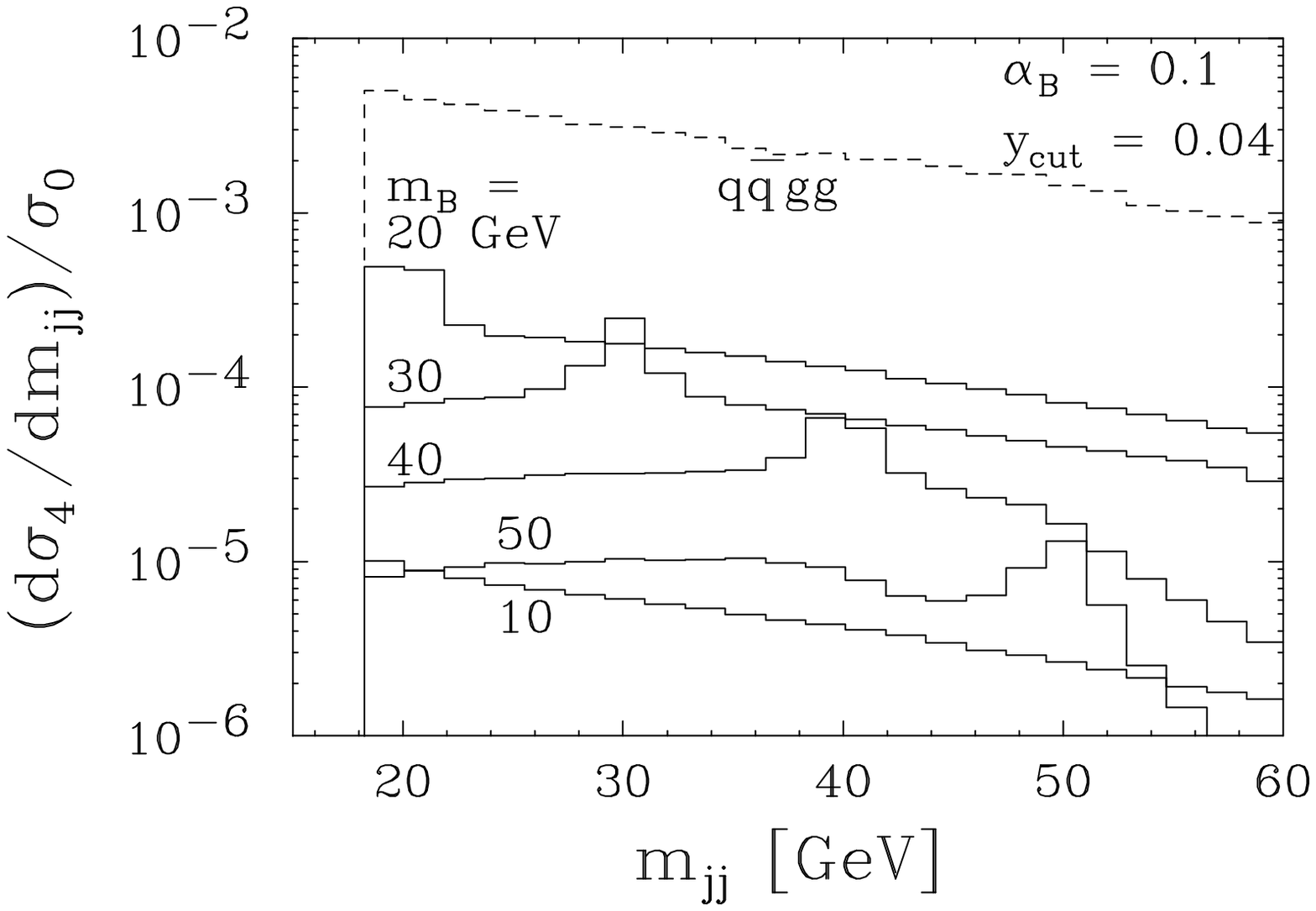,width=2.75in,angle=0}}
\caption{(a) Four-jet cross section as a function of
$y_{{\rm cut}}$ for $\alpha_B=0.1$, normalized to the leading two-jet
cross section. (b) Di-jet invariant mass distribution in four-jet events,
for $\alpha_B=0.1$ and $y_{{\rm cut}}=0.04$, normalized to the leading
two-jet cross section.}
\end{figure}

{\em Di-jet invariant mass peak in $Z\rightarrow 4\,\mbox{jets}$.}
We show the $m_{jj}$ distributions in Fig.~2b for various
values of $m_B$, together with the QCD background. We chose $y_{{\rm cut}} =
0.04$ to optimize the signal for $m_B = 20$~GeV.  It is clear that the
signal is overwhelmed by the background, and hence no practical constraint
exists from the $m_{jj}$ distribution.  Note that existing experimental
searches for dijet invariant mass peaks have required associated
peaks in both pairs of jets (as one would expect, for example, in charged
Higgs production) and thus are irrelevant to our problem.

{\em $\Upsilon(1S)$ Decay.}    The decay of $\Upsilon(1S)$ is another
place to look for the effect of the $\gamma_B$ boson, through its
s-channel contribution to $R_\Upsilon =\Gamma (\Upsilon \rightarrow
\mbox{hadrons})/\Gamma(\Upsilon \rightarrow \mu^+ \mu^-)$.
If we again require that the the resulting shift in the value of
$\alpha_s$ away from the world average to be within two standard
deviations of the value measured in $\Upsilon$ decay, $\alpha_s (m_Z) = 0.108
\pm 0.010$ \cite{PDG}, we exclude the region shown in Figure~1.
One can see that the interesting region of large coupling lies above $\sim
20$~GeV, and thus we do not discuss the region below $m_\Upsilon$ any
further.

\section{Mixing effects}

Clearly, the parameter $c$ must be quite small so that the kinetic mixing
does not conflict with precision electroweak measurements.
Let us begin by defining separate mixing parameters for the $\gamma_B$-photon
and $\gamma_B$-$Z^0$ mixing, namely $c_\gamma$ and $c_Z$.  While
$c_\gamma=c_Z=c$ above the electroweak scale, $c_\gamma$ and $c_Z$ run
differently in the low-energy effective theory below $m_{{\rm top}}$.
Thus, even if $c$ is vanishing at some scale $\Lambda$, $c_\gamma$ and $c_Z$
are renormalized by the one quark-loop diagrams that connect the $\gamma_B$
to either the photon or $Z$, and monotonically increase at lower energies.

The most significant constraints on $c_Z(m_Z)$ are shown in Fig.~3a.
We have considered the effects of the $\gamma_B$-$Z$ mixing on
the following experimental observables: the $Z$ mass, hadronic width,
and forward-backward asymmetries, and the neutral current $\nu N$ and
$e N$ deep inelastic scattering cross sections \cite{carmur}.  We obtained
the strongest constraint from the $Z$ hadronic width.  In addition to the
contributions described earlier, there is an additional contribution
to $Z\rightarrow q\overline{q}$ when $c\neq 0$ due to the $\gamma_B$-$Z$
mixing, which is included in Figure~3a. The hadronic width places the tightest
constraint on $c_Z(m_Z)$, roughly $|c_Z(m_Z)| \lesssim 0.02$.

\begin{figure}
\vspace*{13pt}
\centerline{\psfig{file=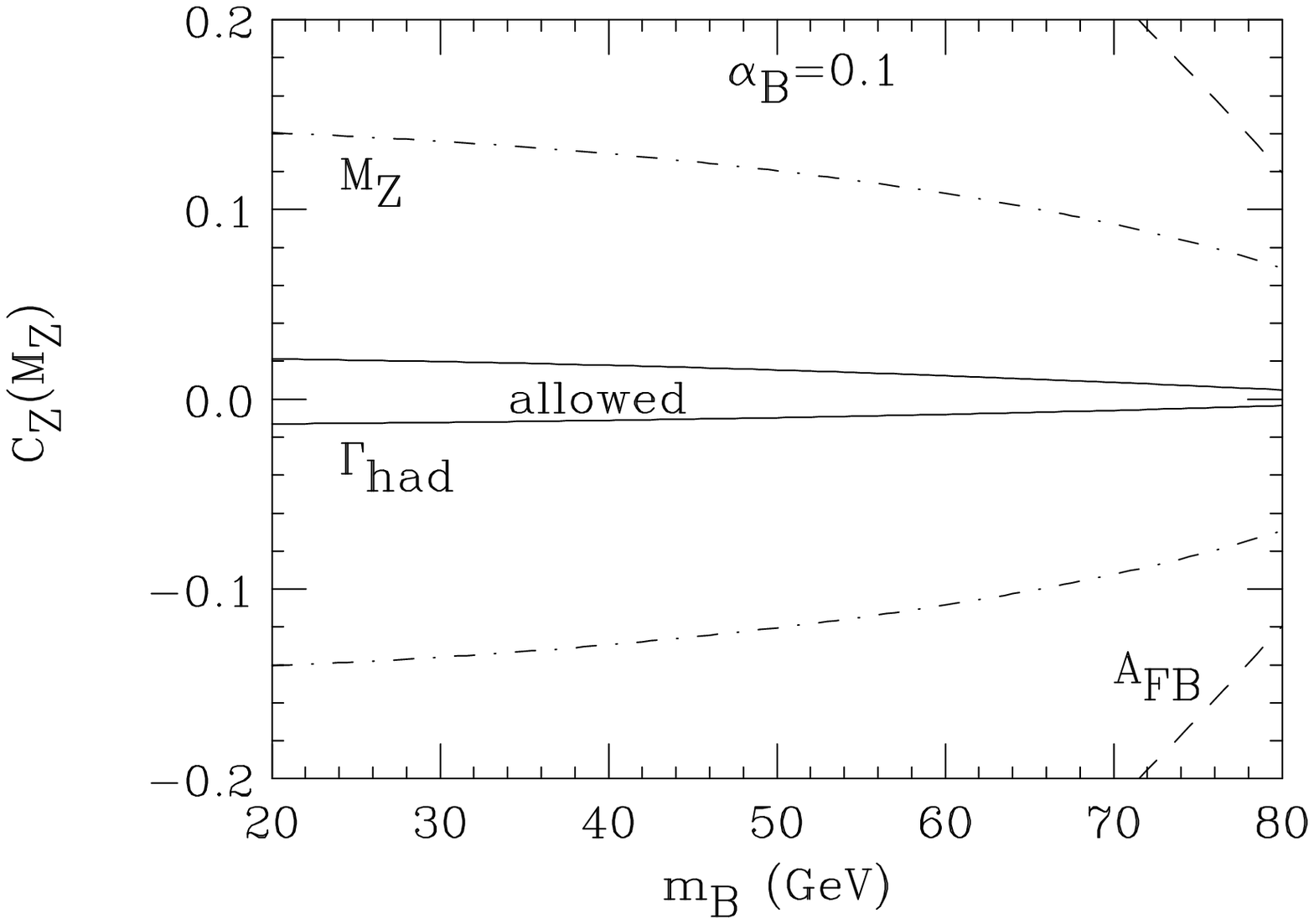,width=2.75in,angle=0}
\psfig{file=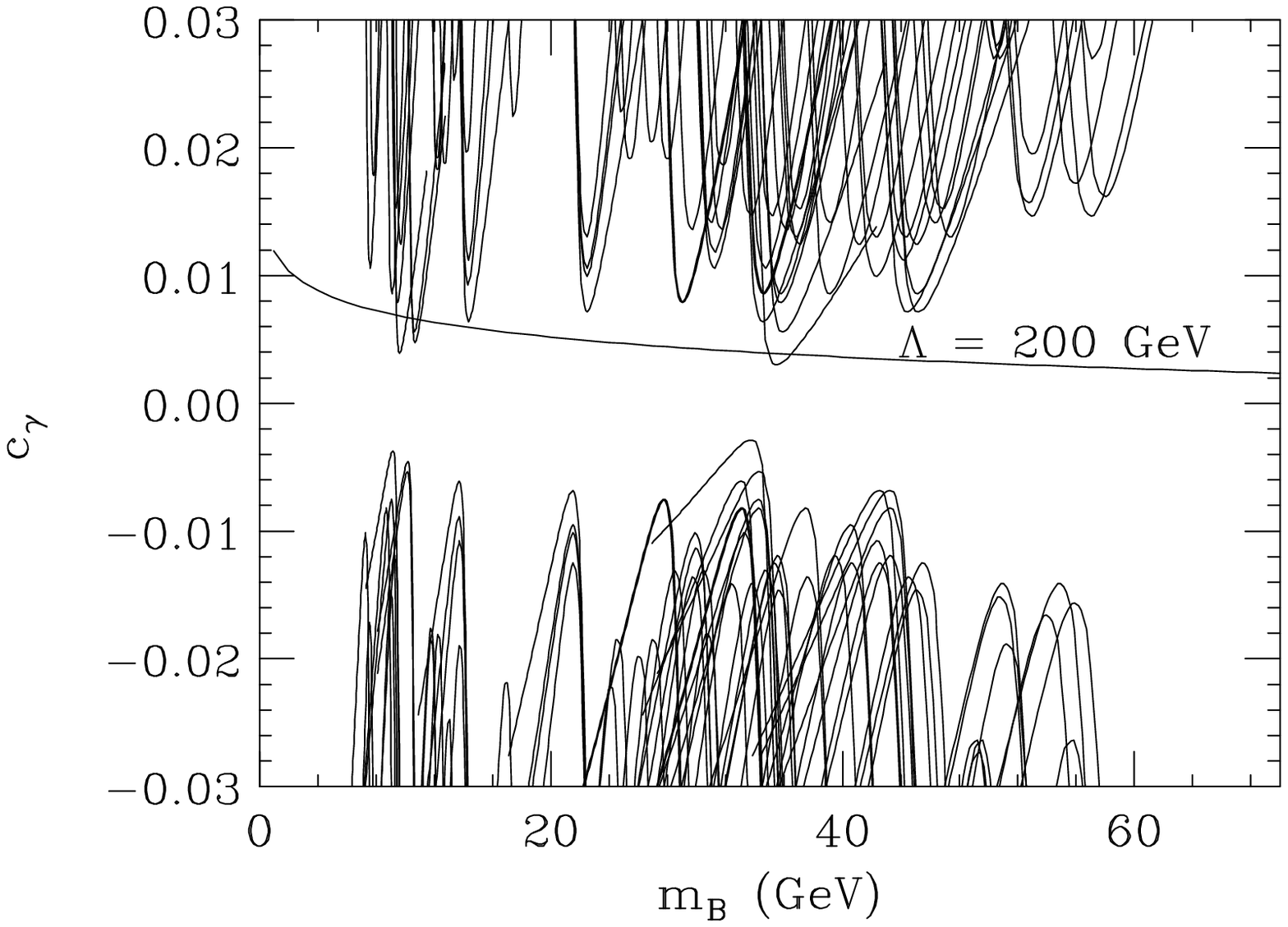,width=2.75in,angle=0}}
\caption{(a) Constraints on $c_Z(m_Z)$ from the two standard deviation
experimental uncertainties in the $Z$ mass, hadronic
width, and $Z\rightarrow b\overline{b}$ forward-backward asymmetry.
(b) Constraints on $c_\gamma(m_B)$ from the two standard deviation
experimental uncertainty in $R$ measured at various
values of $s$ as compiled by the Particle Data Group.
The running of $c_\gamma$ corresponding to $\Lambda=200$ GeV is shown
for comparison.}
\end{figure}

The coupling $c_\gamma$ has its most significant effect on
a different set of observables.  We have considered the effect of
the $\gamma_B$-$\gamma$ mixing on the cross section for
$e^+e^-\rightarrow hadrons$, and on the anomalous
magnetic moments of the electron and muon \cite{carmur}. We obatined the
strongest constraint from the additional contribution to $R$,
the ratio $\sigma(e^+e^-\rightarrow \mbox{hadrons})/
\sigma(e^+e^-\rightarrow \mu^+\mu^-)$. For any $m_B$ of interest, we can
constrain $c_\gamma$ by considering the two standard
deviation uncertainty in the value of $R$ measured at $\sqrt{s}\approx m_B$.
The results are shown in Fig.~3b, based on the cumulative data on $R$ taken
at various values of $\sqrt{s}$ and compiled by the
Particle Data Group \cite{PDG}. Roughly speaking, the allowed region of
Fig.~3b corresponds to $|c_\gamma(m_B)|<0.01$.

Using the  approximate bound $c_Z(m_Z)<0.02$ from above, and assuming
$m_{{\rm top}}\approx 175$ GeV and $\alpha_B=0.1$, we find $c(\Lambda)=0$
for $\Lambda < 1.3$ TeV.  This implies that the scale of new physics lies
at relatively low energies, just above the electroweak scale.

What type of model can naturally satisfy the boundary condition what we
obtained above? Consider a model with the gauge structure
$\mbox{SU(3)}_{C} \times \mbox{SU(2)}_L \times \mbox{U(1)}_Y
\times \mbox{SU(4)}_{H}$, where SU(4)$_H$ is a horizontal symmetry.
In addition to the ordinary three families of the standard
model, $f^i$~$(i=1,2,3)$, we assume there is a fourth family $F$. The
horizontal symmetry acts only on the quarks in the four families, which
together transform as a {\bf 4} under the SU(4)$_H$.  The U(1)$_B$ gauge group
is embedded into SU(4)$_H$ as $diag(1/3,1/3,1/3,-1)$.  The horizontal
symmetry SU(4)$_H$ is broken at a scale $M_H$ down to U(1)$_B$.  It is easy
to see that the kinetic mixing remains vanishing down to the weak
scale: Above $M_{H}$, the mixing is not allowed by gauge invariance
because U(1)$_B$ is embedded into the non-abelian group SU(4)$_{H}$.
Below $M_H$, the particle content satisfies the orthogonality condition
$\mbox{Tr}(BY)=0$ and the mixing parameter $c$ does not run.  The
running begins only after the heaviest particle contributing to
$\mbox{Tr}(BY)$ (i.e. the heaviest fourth generation fermion)
is integrated out of the theory, so that the one-loop diagram
connecting baryon number to hypercharge is nonvanishing.
Since the fourth generation fermions have electroweak scale masses, the
mixing term remains vanishing down to the weak
scale, {\it i.e.},\/ $\Lambda = m_F \sim m_{\rm top}$, and the desired
boundary condition is naturally achieved.

\section{Leptonic Signals}

Finally, we point out that the small kinetic mixing term described above can
provide a possible leptonic signature for our model through the Drell-Yan
production of lepton pairs at hadron colliders. The quantity of interest is
$d\sigma/dM$, the differential cross section as  a function of the lepton pair
invariant mass.  We computed $d\sigma/dM$ in a $p\overline{p}$ collision
at $\sqrt{s}=1.8$ TeV, integrated over the rapidity interval $-1<y<1$,
using  the EHLQ Set II structure functions \cite{ehlq}.  Assuming
$\alpha_B=0.1$ and $c_\gamma(m_B)=c_Z(m_B)=0.01$, we determined the
excess in the total dielectron plus dimuon signal in a bin of size $dM$
surrounding the $\gamma_B$ mass.  The results are shown in Table~1,
for $m_B=30$, $40$, and $50$ GeV. The statistical significance of the signal
assuming integrated luminosities of 1 fb$^{-1}$ and 10 fb$^{-1}$ is also
shown. These results imply that there would be hope of detecting
the $\gamma_B$ boson at the Tevatron with the main injector, and/or an
additional luminosity upgrade.

\begin{table}[h]
\caption{Excess Dielectron plus Dimuon Production at the Tevatron,
with $\alpha_B=0.1$ and $c_\gamma(m_B)=c_Z(m_B)=0.01$.}
\begin{center}
\begin{tabular}{cccc|cc}
\hline
$m_B$ & $dM$ & Background & Excess
      & \multicolumn{2}{c}{statistical significance}\\ \cline{5-6}
(GeV) &(GeV) &   (fb)    & (fb)   & 1~fb$^{-1}$ & 10~fb$^{-1}$\\
\hline\hline
30 & 2 & 3468 & 320 & 5.4~$\sigma$ & 17.2~$\sigma$ \\
40 & 4 & 2798 & 208  & 3.9~$\sigma$ & 12.4~$\sigma$ \\
50 & 4 & 1422  & 112  & 3.0~$\sigma$ & 9.4~$\sigma$ \\
\hline\hline
\end{tabular}
\end{center}
\end{table}

\section{Conclusions.}
We have shown that a new light U(1) gauge boson
$\gamma_B$ coupled to the baryon number evades all existing
experimental constraints in the interesting mass region,
$m_\Upsilon \lesssim m_B \lesssim m_Z$. In this range, the
coupling $\alpha_B$ may be as large as $\sim 0.1$, and
the $\gamma_B$ may have a visible collider signature, even at
existing accelerators.  In particular, the new contribution to
Drell-Yan dilepton  production may yield a detectable signal at the
Fermilab Tevatron after the main injector upgrade.

\section{Acknowledgments}
This work was supported by the Director, Office of Energy Research,
Office of High Energy and Nuclear Physics, Division of High Energy
Physics of the U.S. Department of Energy under Contract DE-AC03-76SF00098.


\end{document}